# Diffractive and Exclusive Production at the Tevatron

M. E. Convery for the CDF Collaboration
*FNAL, Batavia, IL 60510, USA*

We report new results from CDF on single-diffractive dijet and W/Z production, and on exclusive dijet, diphoton, and dilepton production in proton-antiproton collisions at √s=1.96 TeV at the Fermilab Tevatron. The measured exclusive dijet and diphoton cross sections provide a test of theoretical predictions for exclusive Higgs production at the LHC.

## 1. INTRODUCTION

Exclusive Higgs production, in which the event consists of nothing but the leading protons and a Higgs boson, has been proposed as a channel in which to study the properties of the Higgs boson at the Large Hadron Collider (LHC). Although we do not expect to observe exclusive Higgs-boson production at the Tevatron, we can observe similar processes which provide a calibration for theoretical predictions of exclusive Higgs production at the LHC. The CDF measurements of exclusive dijet and diphoton production, examples of such processes, are presented in Sec. 2.

Single diffraction has been studied extensively at the Tevatron in Run I, including diffractive dijet and W/Z-boson production. New results with extended kinematical reach allowed by the larger Run II dataset are presented in Sec. 3.

CDF II includes forward detectors designed for studying diffractive physics. The MiniPlug calorimeters cover the pseudorapidity region $3.5<|\eta|<5.1$. Beam Shower Counters (BSC) surrounding the beampipe in several locations detect particles in the forward region $5.4<|\eta|<7.4$. A spectrometer consisting of three Roman-pot detectors preceded by Tevatron dipoles is used to track diffractive antiprotons which have lost a fraction $0.03<\xi<0.10$ of the beam momentum.

## 2. EXCLUSIVE PRODUCTION

In order to test and calibrate theoretical predictions for exclusive Higgs production at the LHC, CDF has studied the exclusive production of dijets and diphotons, which are also expected to proceed through a colorless hadronic exchange (Fig 1). Exclusive dilepton production via QED two-photon exchange has also been measured. These events consist of nothing but the leading (anti)protons and the object of interest, and their topology is similar to double-pomeron exchange (DPE) in which both the proton and antiproton interact diffractively, escaping intact with a large fraction of the beam momentum, and leaving a large rapidity gap empty of particles in the forward region on both sides.

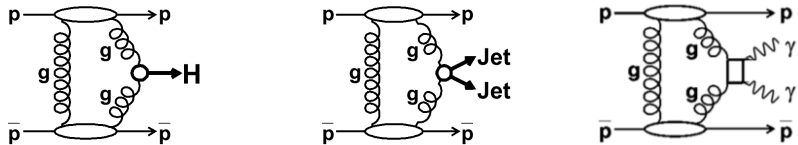

Figure 1: Diagrams for exclusive production of (left) Higgs, (center) dijets, and (right) diphotons.

### 2.1. Exclusive Di-jet Production

There has been a wide range of predictions for the cross section for exclusive Higgs production, and similarly for exclusive dijet production. In Run I, CDF measured DPE dijet production and set a limit on exclusive production [1].

The Run II CDF search [2] starts by requiring a track in the Roman-pot spectrometer, a rapidity gap in the BSC on the proton side, and a pair of jets. Since jets are complicated objects, defined in this case using a cone algorithm, and may have energy outside the cone, the dijet mass fraction defined as the ratio of the dijet mass $M_{jj}$ to the total system mass $M_X$, $R_{jj} = M_{jj}/M_X$, is used to search for exclusive events. The excess at large dijet mass fraction over the inclusive DPE dijet prediction using the diffractive POMWIG Monte Carlo (MC) generator is clearly seen in Fig. 2. Although the shape of this excess is described well by two exclusive dijet MC generators, ExHume and DPEMC, the cross section observed by CDF clearly disfavors DPEMC, as shown in Fig. 2. The ExHume MC is based on a theoretical calculation [3] by Khoze, Martin and Ryskin which has an uncertainty of a factor of three; scaling that prediction down by a factor of three gives a cross section which is consistent with the CDF data (Fig. 2).

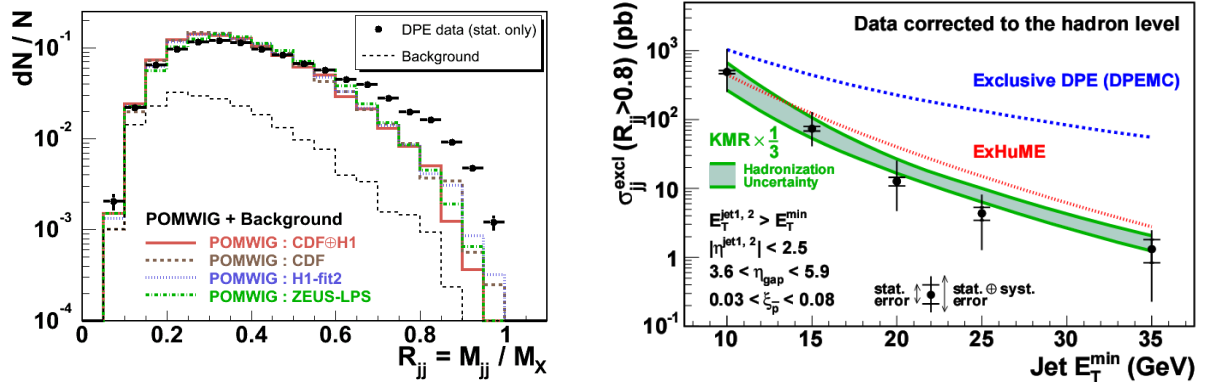

Figure 2: Dijet mass fraction $R_{jj}$ in inclusive DPE dijet data (left). An excess over predictions at large $R_{jj}$ is observed as a signal of exclusive dijet production. The cross section for events with $R_{jj}>0.8$ is compared to predictions (right).

### 2.2. Exclusive Di-lepton and Di-photon Production

Exclusive $e^+e^-$ and exclusive diphoton production have been studied at CDF using a trigger which requires a rapidity gap on both sides and two electromagnetic clusters in the calorimeter. All calorimeters are then required to be otherwise empty, and no tracks (except those associated with the electrons in $e^+e^-$ production) are allowed. CDF finds 16 exclusive $e^+e^-$ candidates, consistent with the expectation from QED. This was the first observation ($5.5\sigma$) of an exclusive two-photon process ($\gamma\gamma\to ee$) in hadron-hadron collisions [4]. Exclusive $\mu^+\mu^-$ studies are in progress.

Three candidate exclusive diphoton events were found in this same 0.5 fb$^{-1}$ data set; two of these events are likely $\gamma\gamma$, and the third is more likely $\pi^0\pi^0$. The probability of observing at least three such events is found to be $1.7\times10^{-4}$, which allows us to put a limit on exclusive diphoton production of 410 fb at 95% CL [5]. The prediction by Khoze, et al. [6] is compatible with this limit. CDF plans to update this measurement with additional available data.

### 3. DIFFRACTIVE PRODUCTION

In hard single diffraction, we examine the partonic structure of the diffractive exchange using high-$p_T$ probes such as jets. The diffractive exchange is defined by the fraction $\xi$ of the antiproton momentum carried away in the exchange (we refer to the exchanged object as a "pomeron"), and the four-momentum-transfer-squared t. The hard process is defined as usual in perturbative QCD by the momentum fraction x of the parton involved in the hard process and the energy scale $Q^2$. The momentum fraction $\xi$ can be determined from the track in the Roman-pot detectors by determining the momentum retained by the antiproton; the Roman pots have good acceptance for $0.03<\xi<0.10$.

The $\xi$ can also be determined from the energy in the rest of the event (summing over calorimeter towers) as

$$\xi_{cal} = \sum_{towers} \frac{E_T}{\sqrt{S}} e^{-\eta} \qquad (1)$$

where it is important to have the MiniPlug calorimeters covering large $|\eta|$. In events with multiple proton-antiproton interactions, $\xi_{cal}$ contains energy from more than just the diffractive interaction, and allows us to remove events with an overlapping interaction by requiring $\xi_{cal}<0.10$, and to estimate the remaining overlap background at smaller $\xi_{cal}$.

### 3.1. Diffractive Di-jet Production

The diffractive structure function is a function of x, $Q^2$, $\xi$, and t, and was determined as a function of x in different ranges of $\xi$ by CDF in Run I in diffractive dijet production, simply by taking the ratio of the diffractive to non-diffractive dijet cross section, which is to good approximation the ratio of the diffractive to the known proton structure function. In Run II, CDF was able to confirm this result and to look at the structure function in ranges of $Q^2$, going as high as $10^4$ GeV$^2$, and found no appreciable $Q^2$ dependence [7], suggesting that the pomeron evolves similarly to a proton. CDF has also looked at the diffractive dijet cross section as a function of t in ranges of $Q^2$ up to 4500 GeV$^2$, and finds no dependence of the shape of the t distribution on $Q^2$ [7].

### 3.2. Diffractive W- and Z-Boson Production

The study of diffractive W- and Z-boson production helps to determine the quark content of the pomeron since, to leading order, the W/Z is produced through a quark in the pomeron. Run I studies used rapidity gaps to identify diffractive events. CDF found the fraction of W events which are diffractive to be [1.15±0.51(stat)±0.20(syst)]% [8]. By combining this result with diffractive dijet production, which probes both quarks and gluons in the pomeron, and also a measurement of diffractive b-quark production, CDF determined the gluon fraction of the pomeron to be $[54^{+16}_{-14}]$% [9]. D0 reported the fraction of events with a rapidity gap, uncorrected for the gap survival probability <S>, to be $[0.89^{+0.19}_{-0.17}]$% of W's and $[1.44^{+0.61}_{-0.52}]$% of Z's [10], with <S> ~ 20% to 100% for models considered.

In the CDF Run II measurement, the Roman pots provide an accurate $\xi$ measurement, as well as eliminating the problem of the gap survival probability. As in Eq. 1, we can calculate $\xi$ from the energy in the calorimeter (Fig. 3); however, in the case of W→$l\nu$ production, $\xi_{cal}$ is missing the energy from the neutrino. The difference between the true $\xi$ measured in the Roman-pot detectors, $\xi_{RP}$, and $\xi_{cal}$ allows us to determine the neutrino kinematics (in particular, $\eta_\nu$, while the transverse energy of the neutrino is taken to be the missing $E_T$ in the event)

$$\xi_{RP} - \xi_{cal} = \frac{\not{E}_T}{\sqrt{S}} e^{-\eta_\nu} \qquad (2)$$

and therefore, the full W kinematics. For example, the reconstructed W mass is shown in Fig. 3. CDF hopes to use this to determine the diffractive structure function from W production. Since we expect $\xi_{cal} < \xi_{RP}$, we impose this requirement in order to remove events with multiple proton-antiproton interactions, and can then be sure that the W was produced in the diffractive interaction. We further cut on the reconstructed W mass $50<M_W<120$ GeV/c$^2$ in order to remove possible misreconstructed events. We then find a fraction of W's which are produced diffractively (in the region of good Roman-pot $\xi$, t acceptance) to be $R_W(0.03<\xi<0.10, |t|<1)=[0.97\pm0.05(stat)\pm0.11(syst)]\%$, consistent with the Run I CDF result when corrected for the $\xi$ and t range measured. The W $p_T$ distribution is also shown in Fig. 3.

We find 37 diffractive Z candidates with $\xi_{cal}<0.10$ and estimate that 11 of these are overlap events, based on the non-diffractive $\xi_{cal}$ distribution, as was done in the diffractive dijet study. This gives a fraction of Z's which are diffractive of $R_Z(0.03<\xi<0.10, |t|<1)=[0.85\pm0.20(stat)\pm0.11(syst)]\%$. This is the full data sample with the Roman pots, 0.6 fb$^{-1}$.

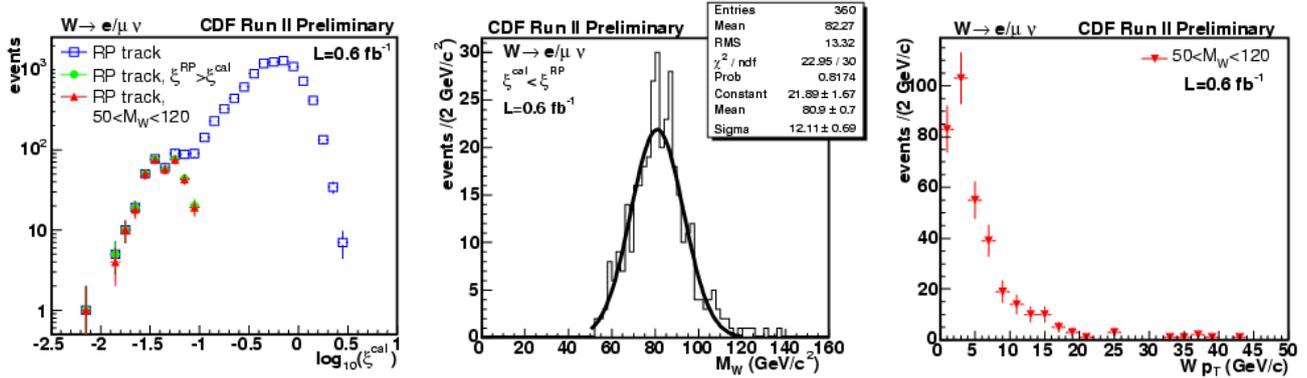

Figure 3: Calorimeter $\xi$ distribution in W events with a Roman-pot track (left). Since the calorimeter is missing the neutrino energy, we expect $\xi_{cal} < \xi_{RP}$ for events with no overlap interaction. The difference $\xi_{RP} - \xi_{cal}$ allows us to determine the W mass in diffractive events (center). The $p_T$ distribution of diffractively-produced W's (right).

## 4. CONCLUSIONS

Measurements of exclusive production at CDF, including the observation of exclusive dijet production and setting limits on exclusive diphoton production, have provided a calibration essential for predictions for exclusive Higgs production at the LHC. The diffractive structure function and t distribution from dijet production have been measured in ranges in Q$^2$. The measurement of diffractive W- and Z-boson production using Roman pots is found to confirm the CDF Run I rapidity-gap result.